\documentclass{appolb}
\usepackage{epsfig}

\newcommand{\mb}[1]{ { \mbox{\boldmath{$#1$}}}  }
\newcommand{\mbs}[1]{ {\scriptsize \mbox{\boldmath{$#1$}}}  }
\newcommand{\sub}[2]{ \mbox{$#1$}_{\mbox{\scriptsize\boldmath{$#2$}}}}

\begin{document}
\title{Disorder Induced Fluctuations of the Pairing  Parameter in
P--Wave Superconductors
\thanks{Presented at XII International School of Modern Physics on Phase 
Transitions and Critical Phenomena, L\c{a}dek Zdr\'oj 2001}%
}
\author{Grzegorz Litak
\address{Department of Mechanics, Technical University of Lublin,
Nabystrzycka~36, 
20-618 Lublin, Poland
}
}
\maketitle
\begin{abstract}
 We study the effect of site diagonal disorder
on the
pairing amplitude
by a perturbation
method. Using  an extended Hubbard model with the intersite
attraction we analyze
 fluctuations of order parameter in presence of
 non-magnetic disorder and  discuss the
instability of
various  solutions with  $p$--wave paring.
\end{abstract}

\section{Introduction}

In case of an anisotropic superconductor non-magnetic disorder
leads eventually to pair breaking effect via Abrikosov--Gorkov formula
\cite{Abr59,Mak69,Har01,Mar99,Lit00}, 
like for magnetic impurities, for
any value of a coherence length $\xi$.
By analogy
to an isotropic $s$--wave pairing, for superconductors
with an enough large coherence length
$\xi$, the amplitude of paring
potential $|\Delta(ij)|$ tends
be the same for all
bonds but 
disorder induces the effect of
fluctuations in the amplitude leading, in this way, to
destruction of pairing.
Here we will calculate
fluctuating potentials standard  deviations 
$< \delta |\Delta_{ij}|^2>$
and $< \varepsilon_i^2 >$
Their ratio $\Gamma=< \delta |\Delta_{ij}|^2>/
< \varepsilon_i^2 >$ 
will be a criterion of pairing potential
fluctuations \cite{Gyo97}.

\begin{figure}[htb]
\leavevmode
\vspace{-1.0cm}

\hspace*{0.5cm}
\epsfxsize=4.5cm
\epsffile{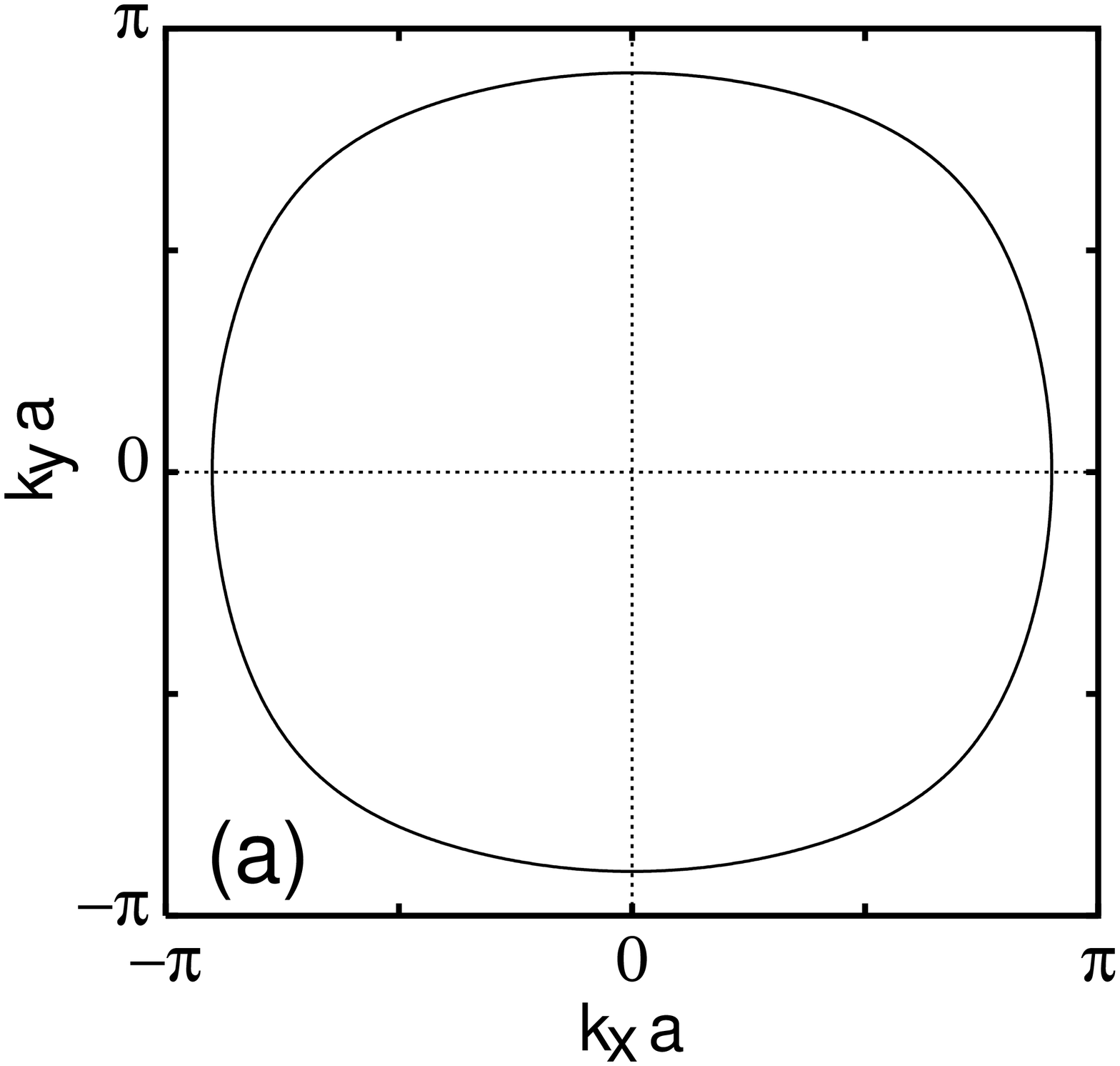}
\hspace*{0.5cm}
\epsfxsize=4.5cm
\epsffile{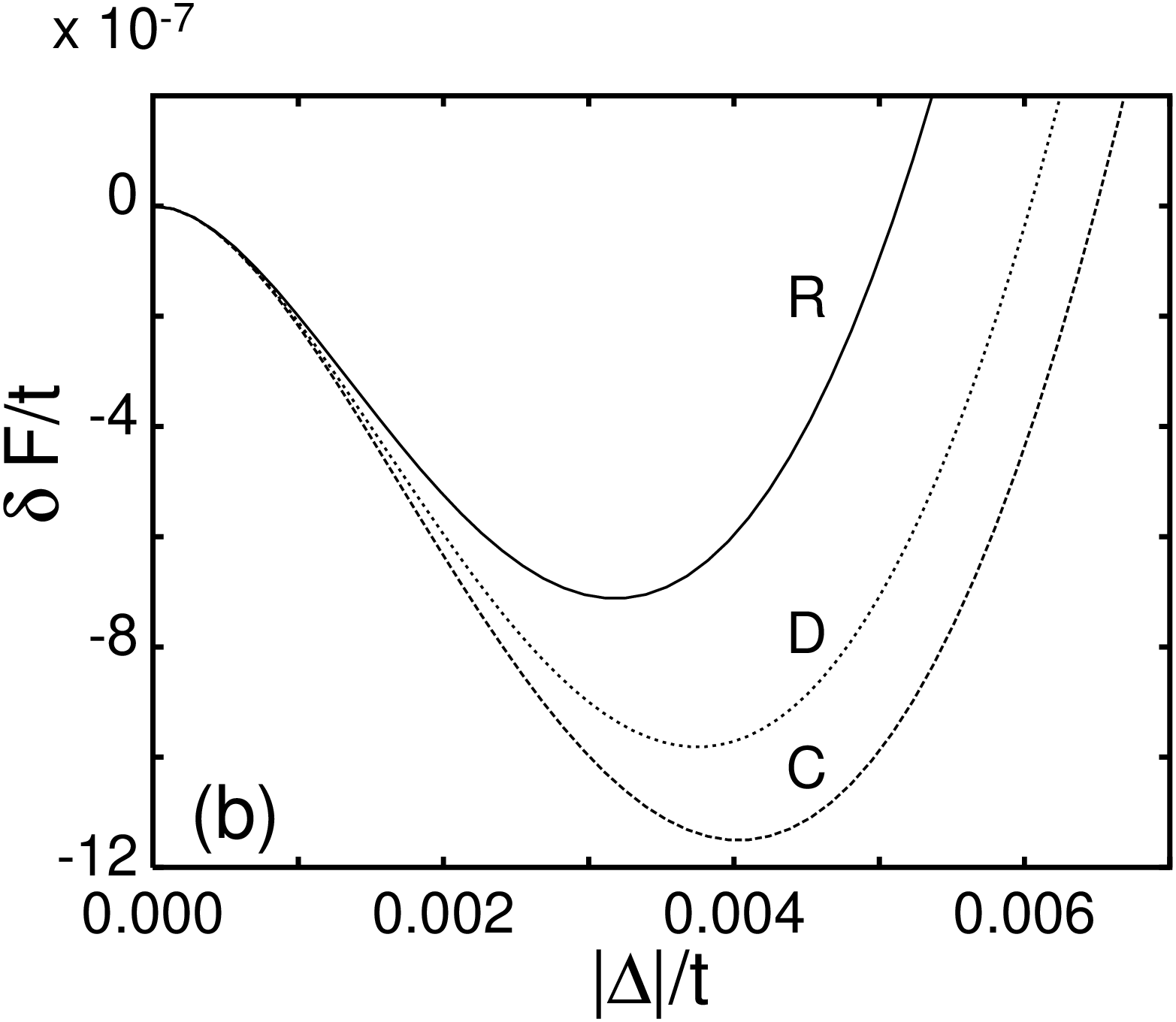}
\vspace{0.0cm}
\caption{(a) The Fermi surface and (b) the Free energy $\delta F$ as a
function of pairing potential 
$|\Delta|$ ($|\Delta|=
\sqrt{\Delta_x^2+\Delta_y^2}$) for different solutions: dipole (D) real
(R)
and complex (C) order parameters, respectively; for
$\gamma$
band of Sr$_2$RuO$_4$ electronic structure ($t'=0.45t$, $n/2=0.66$). The
intersite
attraction W=-0.446t, temperature $T=0$}
\end{figure}

\section{P--wave solutions for a clean system}

We start a single band, extended, Hubbard model with effective
nearest neighbour intersite
attraction $W_{ij}$ ($W_{ii} =0$) \cite{Mic88}. 
Taking the Fourier transform of the Green function for a clean system
 we can write its equation of motion: 
\begin{equation}
\left[ \begin{array}{c} (\omega-\sub{\epsilon}{k}  +\mu) \mb 1~~~
\sub{\mb \Delta}{k}
 \\
\sub{\mb \Delta}{k}^*~~~   (\omega
+\sub{\epsilon}{ k} -\mu) \mb 1   \end{array}      \right] {\mb G}^0(\mb
k; \omega)={\mb 1},
\label{eq5}
\end{equation}
where $\sub{\mb \Delta}{k}= \mb \Delta_{x} \sin k_x +\mb 
\Delta_{y} \sin k_y$ defines
 $2\times 2$
matrix of a pairing potential.
For a clean system $t_{ij}$ can be expressed
in
$\mb k$-space: 
$\sub{\epsilon}{k}= \sum_j t_{ij} {\rm e}^{-\imath \mbs R_{ij} \mbs k}$ $
= -2 t (\cos k_x +
\cos k_y)-4t'\cos k_x \cos k_y$,
where $t$ represents the nearest neighbour site amplitude of electron
hopping, while $t'$ corresponds to next nearest neighbour
one. 
For the assumed solution \cite{Lit01c}
$\sub{\mb \Delta}{k}=
 \imath \hat{\sigma}_y\hat{{\mathbf \sigma}}\cdot{\bf d}({\bf k})$,
with ${\bf d}({\bf k})=(0,0,d^z({\bf k}))$ 
and
 $d^z({\bf k})=\sub{\Delta}{k}$.   
The corresponding free energy $F$ for a finite temperature $T$ can be
found from the
standard formula:
\begin{equation}
\label{eq12}
F=\sum_{\mbs k} \left[ -(n-1) (\sub{\epsilon}{k}-\mu)  -2k_BT~ {\rm ln}
\left( 2
\cosh
\frac{ \sub{E}{k} }{2k_BT} \right) -\frac{|\sub{\Delta}{k}(T)|^2
}{W} \right],
\end{equation}
where $\sub{E}{k}$ denotes the eigenvalue. 
To perform numerical calculations we have fitted our one band
system
parameters
to the realistic $\gamma$ band structure of Sr$_2$RuO$_4$ 
\cite{Lit00,Lit01a,Mac96,Agt97} 
Fig. 1a presents the corresponding Fermi surface.
For the above assumptions we have found
three solutions with
$p$--wave
pairing.
Namely, depending on relative values $\Delta_x$ and
$\Delta_y$, the dipole one ($\Delta_x \neq 0$ and  $\Delta_y = 0$),
the real one ($\Delta_x=\Delta_y$) and the complex one ($\Delta_x=\imath
\Delta_y$).
They correspond  to minima of free energy curves ($\delta F(|\Delta|)=
F(|\Delta|)-F(0)$)
in
Fig. 1b, denoted by D, R and  C, respectively.  
The interaction parameter used in calculations (W=-0.446t) were chosen to
give $T_c=1.5$K as for clean  Sr$_2$RuO$_4$. 
One can easily see that the complex solution reaches the global minimum of
free
energy $F$.

\section{ Fluctuations of pairing potential.}  

In this section we investigate the stability of superconducting $p$--wave
states in presence of a weak disorder.
Here, we apply the same strategy as in \cite{Gyo97} and we
treat random site energies $\varepsilon_i$ as
perturbations. To proceed we write the Dyson equation for a Green
function $\mb G(i,j;\omega)$ evaluated at a frequency
$\omega$:

\begin{equation}
\label{eq13}
{\mb G}(i,j;\omega) = {\mb G}^0(i,j;\omega) + \sum_l
{\mb G}^0(i,j;\omega) {\mb V}_l {\mb G}(l,j;\omega),~
\end{equation}
where $\mb V_l= \varepsilon_l {\mb \sigma}_3 $ is the impurity potential
matrix.

Following Eq. (\ref{eq13}) we  express
quantity order parameter $\Delta_{ij}$, in the lowest
order
of  $\varepsilon_i$ perturbations by means of disordered Green function
and calculate the mean square deviation  of the paring
$\Delta_{ij}$ along the bond of nearest neighbour sites $i$ and $j$.
For site independent energies $\varepsilon_i$:

\begin{equation}
< \delta |\Delta_{ij}|^2>
= \Gamma_{ij} < \varepsilon_i^2 >.
\label{eq17}  
\end{equation}

Finally, we calculate the coefficient $\Gamma_{ij}$ 
\cite{Gyo97}:

\begin{equation}
\label{eq18}
\Gamma_{ij} =  \frac{1}{N} \sum_{\mb q}
\left|\frac{W_{ij}}{2N}
\sum_{\mb k}
\frac {\sub{\Delta}{k} \tilde{\epsilon}_{\mb k} + \sub{\Delta}{k} 
\tilde{\epsilon}_{\mb k - \mb q}} {(
E_{\mb k}+ E_{\mb k - \mb
q})
 E_{\mb k} E_{\mb k - \mb q}}{\rm e}^{\imath(\mb R_i -\mb R_j)
\mb k}
\right|^2~,
\end{equation}

Having found the pairing potentials (Fig. 1b, Eq. \ref{eq5})
for the clean
system, we calculated
$\Gamma$ (Eq. \ref{eq18}) for all three
solutions we obtained. 
In all cases $\Gamma$ has a  very small value (of order $\sim 10^{-8}$).
This implies that 
fluctuations of $\Delta_{ij}$ are relatively small in this system.
Interestingly,
a real type  solution is characterized by  the smallest fluctuations
($\Gamma=5.15
\cdot 10^{-9}$) while $\Gamma=13.93 \cdot 10^{-9}$ for the dipole solution
and $\Gamma=8.12 \cdot 10^{-9}$ for the complex one. This could 
mean that  the real solution is favoured by disorder.

\section{Conclusions and Discussion}

We have analyzed the effect of a weak disorder on a $p$--wave
superconductor
in context of newly discovered superconductor  Sr$_2$RuO$_4$
\cite{Agt97,Mae01}.
 Unfortunately
the order parameter structure in this compound is still unknown
\cite{Mae01}.
Fitting our system parameters to its $\gamma$ band structure we have asked
about
the stability of various solution in presence of disorder.
We have found three solutions with the same critical temperature
$T_C=1.5$K: dipole and real solutions with line nodes in the gap and
complex
one with a finite gap in any direction.
Note that, all these solutions have the same critical temperature
$T_C$ and the Abrikosov--Gorkov formula
\cite{Mak69,Mar99} does not differentiate
any of them.    
Our preliminary results at $T=0$ indicate that the complex type of
solution has
the global minimum of free energy $F$ but the real one is favoured
by 
disorder. 
That result was obtained in one band model in the lowest order
of perturbation
method and should be confirmed  by a more sophisticated method like 
Coherent Potential Approximation \cite{Mar99,Lit00}
considering  more realistic three bands structure of Sr$_2$RuO$_4$.

\vspace{1cm}

\noindent {\bf Acknowledgements:}
This work has been partially supported by KBN grant No. 5P03B00221 and
 the
Royal Society Joint Project.
The author would like to thank Profs. K.I. Wysoki\'nski, B.L. Gy\"{o}rffy
and Dr. J.F. Annett
for discussions.

\end{document}